\documentstyle[11pt,psfig,amssymb]{article}
\newcommand{\Ht}{\hat{t}}
\newcommand{\HG}{\hat{G}}
\newcommand{\Hg}{\hat{g}}
\newcommand{\HI}{\hat{I}}
\newcommand{\base}{}

\begin{document}
\renewcommand{\thepage}{ }
\begin{titlepage}
\title{
{\center \bf Detection of non separable correlations\\
with dc transport:\\
a new type of Aharonov-Bohm effect}}
\author{
R. M\'elin\thanks{melin@polycnrs-gre.fr}
{}\\
{}\\
{Centre de Recherches sur les Tr\`es Basses
Temp\'eratures (CRTBT)\thanks{U.P.R. 5001 du CNRS,
Laboratoire conventionn\'e avec l'Universit\'e Joseph Fourier
}}\\
{CNRS BP 166X, 38042 Grenoble Cedex, France}\\
}
\date{\today}
\maketitle
\begin{abstract}
\normalsize
I construct a minimal formalism to describe transport
of non separable correlations arising when
a superconductor is in contact with several
electrodes (being ferromagnetic or normal metal).
Transport
theory is expressed in terms of effective
single site
Green's functions. 
Part of the circuit is
decimated exactly by renormalizing only one
physical parameter (the density of states).
I show that
the physical current is obtained by acting on the
infinite series of Feynman diagrams
with an operator counting the charge
of a given diagram. 
I use this method to
propose a new Aharonov-Bohm experiment intended to
detect non separable
correlations. In this experiment, spin--$\sigma$ electrons
are forced to travel around an Aharonov-Bohm loop
in the presence of an applied magnetic flux, while
spin--($- \sigma$) electrons are not directly coupled to
the magnetic flux. It is shown that the spin--($-\sigma$)
current oscillates as a function of
the flux to which
spin--$\sigma$ electrons are coupled.
It is predicted that the effect exists both
with high and low transparency contacts, and that
the electrodes can be normal metal in which
phase coherence can propagate over large distances.
Three other aspects of the problem are 
investigated: (i) The dependence of the superconducting
gap on spin polarization of ferromagnetic
electrodes connected to the superconductor;
(ii) The detection of linear superposition of
correlated pairs of electrons;
(iii) The proximity effect version of
cryptomagnetism.
\base
\end{abstract}
\end{titlepage}
\newpage
\renewcommand{\thepage}{\arabic{page}}
\setcounter{page}{1}
\base

\newpage

\section{Introduction}

There are two important challenges in
phase coherent transport phenomena.
One is to fabricate quantum bits for quantum
computing~\cite{Schor,Grover,Steane,Nakamura,Loss,Hekking}.
The other is to make an Einstein, Podolsky, Rosen (EPR)
experiment, that would be able
to probe Bell inequalities with electrons~\cite{Bell}.
Bell inequalities have already been tested experimentally with
photons~\cite{Aspect,Kwiat}, being massless particles.
EPR experiments constitute one of the ultimate
tests of quantum mechanics, which is a non
local theory without hidden variables (at least
with photons). At the present stage, there exists
no clear proposal of what should be an EPR experiment
with electrons. It is not even known whether
it is possible to propose a realistic situation
in which a transport experiment could detect
the presence/absence of hidden variables
with electrons. However, electrons carry an
electric charge while photons do not.
One can take advantage of this, and propose
experiments with electrons that would
not be possible with photons.

It is already well established that
the basic condition required to design an EPR
experiment with electrons is to have a source
of correlated pairs of electrons~\cite{Loss,Martin}.
Several dc transport experiments
have been proposed recently~\cite{Feinberg,Falci},
and a new proposal is made here. From a technical
point of view, I concentrate on the simplest
non perturbative formulation of transport theory,
which relies on the use of effective single
site Green's functions.
By non perturbative,
I mean that transport theory is exact to
all orders in the tunnel amplitude. 
As a consequence,
I can solve the physics of high transparency contacts.
The interesting aspect of the formalism is that part
of the circuit can be decimated exactly.
Only one
physical parameter is renormalized (the
density of states). I show how to cure a more technical
problem, related to the fact that all possible Feynman
diagrams are present in the theory and have exactly
the same value. The solution is to act on the
infinite series of Feynman diagrams with an operator counting
the charge of a given
diagram.

\medskip

I apply the method to discuss
four relevant physical questions:
\begin{itemize}
\item[(i)] What is the thermodynamics of a superconductor
in the presence of correlated pairs of electrons~?
I show that the strength of superconductivity depends
on whether there exists or not non separable correlations
in the vicinity of the superconductor. This may have
direct implications for experiments.
\item[(ii)] It is possible to fabricate linear superpositions
of correlated pairs of electrons. How should one interpret
transport of such states~? I show that transport
theory is related to quantum measurement {\sl via}
projections of the linear superposition onto
Cooper pair objects.
\item[(iii)] Is there any way to manipulate one electron
making the correlated pair and look at the response of
the other electron~? I show that the answer is
{\sl yes} and propose a new type of Aharonov-Bohm
effect that can be used to probe
non separable correlations. 
Generically, there is no way to manipulate one
electron without avoiding a response of the
other electron.
\item[(iv)] Is there a specific physics
associated to Cooper pair penetration in
ferromagnets in the presence of domain
walls ?
A simple
model is solved, which shows that superconducting
correlations can propagate along ferromagnetic
domain walls. This may apply to recent experiments
in ferromagnet~-- superconductor
heterostructures~\cite{Giroud,Petrashov,Chandra},
and constitutes the proximity version
of cryptomagnetism.
\end{itemize}
Regarding the third question, it is worth recalling that
in ordinary mesoscopic physics Aharonov-Bohm
experiments~\cite{AB,Buttiker}, electrons are forced
to travel through a metallic ring in the presence
of a magnetic flux $\phi$ inside the hole
of the ring. The current oscillates periodically
as a function of $\phi$, with a period equal
to the flux quantum $\phi_0=h/e$~\cite{Manip}.
This is because the system behaves like an interferometer
in which the phase of the electronic wave function
is related to the circulation of the vector potential.
These oscillations occur in a variety
of situations. For instance, 
the dual situation in which a magnetic flux
is forced to
travel around a magnetic charge is known
as the Aharonov-Casher effect~\cite{Casher}.
It is also very often that excitations
of correlated states of matter do not carry a charge $e$.
For instance, Cooper pairs carry a charge $2e$.
The associated flux quantum is $h / (2e)$~\cite{Tinkham}.
Now when
ferromagnetic or normal metal
electrodes are connected to a superconductor,
it is possible to fabricate correlated pairs of electrons
of the type $c_{\alpha,\uparrow}^+ c_{\beta,\downarrow}^+
|0\rangle$ in which the spin-up and spin-down electrons making
the Cooper pair reside in different electrodes. I
propose here a situation in which the spin--$\sigma$
electron making the Cooper pair
is forced to couple to a vector potential (by propagating
on a loop in the presence of a magnetic flux) while
the spin--($-\sigma$) electron is not directly
coupled to the magnetic flux (it propagates in an 
ordinary ferromagnet without any hole). As I show,
{\sl the spin--($-\sigma$) current oscillates as a function of the
flux coupled to spin-$\sigma$ electron}. This new type
of Aharonov-Bohm effect can be viewed as a direct
consequence of non separable correlations, and may be tested
in future experiments.

The article is organized as follows.
As a warm up exercise,
I first calculate the transport formula
associated to non local Andreev reflections in
section~\ref{sec:1}. I introduce the effective single site
formalism in section~\ref{sec:forma}, and provide
a discussion of
higher order Feynman diagrams.
Transport of linear superpositions of correlated
pairs of electrons
is derived in section~\ref{sec:ent}.
The Aharonov-Bohm experiment is discussed
in section~\ref{sec:AB}. I present
in section~\ref{sec:domain}
a simple model for superconducting propagation along a 
domain wall, which appears to be a problem closely
related to non separable correlations.
Final remarks
are given in section~\ref{sec:conclusion}.

\section{Warming up: transport of non local
Cooper pairs}
\label{sec:1}
\begin{figure}
\centerline{\psfig{file=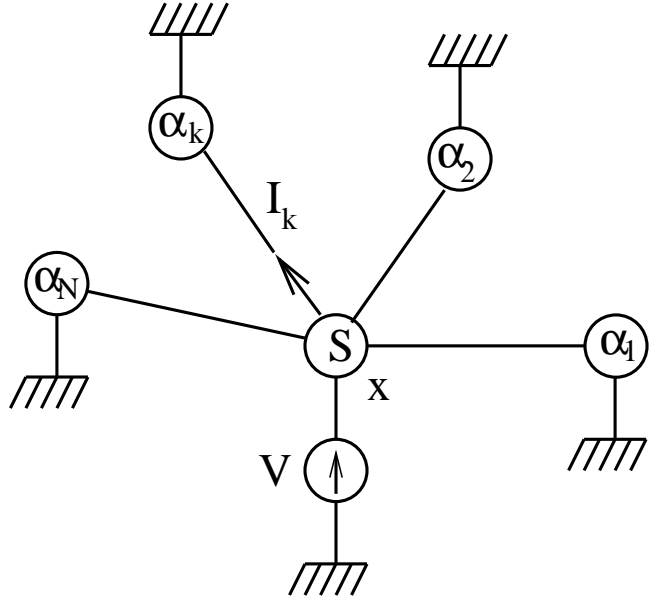,height=8cm}}
\caption{Schematic representation of the model
considered in section~\ref{sec:1} in which a
superconductor (site $x$) is connected to $N$
ferromagnetic electrodes (sites $\alpha_1$, ... ,
$\alpha_N$). A voltage $V$ is applied on the
superconductor.
\base
}
\label{fig:single}
\end{figure}

I start by considering the model
presented on Fig.~\ref{fig:single} in which a
superconductor is connected to $N$
ferromagnetic electrodes.  I first
determine the superdoncuting gap in a
self-consistent way, and next
calculate the current
$I_n$ flowing into one of the ferromagnetic electrodes.

\subsection{Effective single site Green's functions}

Let us first describe the superconductor and
ferromagnetic electrodes 
on the basis of effective single site Green's functions.
The Nambu representation of the 
superconductor Green's function takes the form~\cite{Cuevas}
$\Hg_{x,x}^{A,R}(\omega) = f \HI + g \hat{\sigma}^x$,
with
$$
g(\omega \pm i \eta) = \pi \rho_N \frac{ - \omega \pm
i \eta}{ \sqrt{ \Delta^2 - (\omega \pm i \eta)^2}}
\mbox{, and }
f(\omega \pm i \eta) = \pi \rho_N \frac{ \Delta
}{ \sqrt{ \Delta^2 - (\omega \pm i \eta)^2}}
,
$$
where $\rho_N$ is the normal density of states.
The Nambu representation of the effective single site ferromagnetic
Green's function reads
$
\Hg^{A,R}_{\alpha_k,\alpha_k} = \pm i \pi \left\{
\rho_{k,\uparrow} [ \HI + \hat{\sigma}^z] / 2 +
\rho_{k,\downarrow} [ \HI - \hat{\sigma}^z] / 2 \right\}$,
with $\rho_{k,\sigma}$ the spin--$\sigma$
density of states in the ferromagnetic electrode $\alpha_k$.
The Nambu representation of the hopping matrix
element is
$
\Ht_{x,\alpha_k} = t_{x,\alpha_k} \hat{\sigma}^z
$.
Using effective single site Green's functions 
means that the superconductor is viewed as
as being ``zero dimensional'' . This is a valid
assumption if
the distance between the contacts is
smaller than the coherence length.
This is precisely the condition under which
non local Andreev reflections occur~\cite{Falci}.
Therefore, representing the superconductor by
effective single site Green's functions is a self
consistent assumption.

\subsection{Self consistent
determination of the superconducting gap}
\label{sec:gap-self}
To determine the superconducting gap, we need
to calculate the Gorkov function.
First, we solve the Dyson equation
\begin{equation}
\label{eq:Dyson}
\hat{G}_{x,x}^{R,A}= \left[ \hat{I} - 
\sum_{k=1}^N
\hat{g}_{x,x}^{R,A}
\hat{t}_{x,\alpha_k}
\hat{g}_{\alpha_k,\alpha_k}^{R,A}
\hat{t}_{\alpha_k,x} \right]^{-1}
\hat{g}_{x,x}^{R,A}
.
\end{equation}
The relevant parameters appear to be
the spectral line-width associated to spin-$\sigma$
electrons: $\Gamma_\sigma = \sum_{k=1}^N
\Gamma_{k,\sigma}$, with
$\Gamma_{k,\sigma} = (t_{\alpha_k,x})^2
\rho_{k,\sigma}$. Solving Eq.~\ref{eq:Dyson} leads to
\begin{equation}
\label{eq:solu-G}
\hat{G}_{x,x}^A = \frac{1}{\cal D}
\left\{ \left[ g + i \pi (f^2 - g^2) \right]
\hat{I} + f \hat{\sigma}^x \right\}
,
\end{equation}
with ${\cal D} = 1 - i \pi g(\Gamma_\uparrow
+ \Gamma_\downarrow) + \pi^2 (f^2 - g^2)
\Gamma_\uparrow \Gamma_\downarrow$.
To calculate the superconducting order parameter, we need
to
solve the Dyson-Keldysh equation
$\hat{G}^{+ -} =
(\hat{I} + \hat{G}^R \otimes \hat{W} )
\otimes \hat{g}^{+ -} \otimes ( \hat{I}
+ \hat{W} \otimes \hat{G}^A )$, where
the convolution includes a sum over the labels
$x$ and $\alpha_k$.
Noting $X_\sigma = (1 - i \pi g \Gamma_{- \sigma})/{\cal D}$,
$Y_\sigma = i \pi f \Gamma_\sigma / {\cal D}$,
and using Eq.~\ref{eq:solu-G},
we obtain the exact expression of the Nambu component
of the Keldysh Green's function:
\begin{eqnarray}
\label{eq:solu-Gpm}
{}\left[ G_{x,x}^{+ -} \right]_{2,1} && =
2 i \pi n_F(\omega) \times \\
\nonumber
&{}&\left\{ \rho_g \left( X_\uparrow \overline{Y}_{\uparrow}
+ Y_{\downarrow} \overline{X}_{\downarrow} \right)
+ \rho_f \left( X_\uparrow
\overline{X}_\downarrow + Y_\downarrow
\overline{Y}_\uparrow \right) \right.\\
\nonumber
&{}&\left. + \frac{1}{\pi^2 \Gamma^\downarrow} 
\overline{Y}^\downarrow \left( X^\uparrow -1 \right)
+ \frac{1}{\pi^2 \Gamma^\uparrow} 
Y^\uparrow \left( \overline{X}^\downarrow -1 \right)
\right\}
,
\end{eqnarray}
where $n_F(\omega)$ is the Fermi distribution,
and I used the notation $\hat{\rho}
= \rho_g \hat{I} + \rho_f \hat{\sigma}^x
= \mbox{Im}[\hat{g^A}]/\pi$.
The superconducting gap is
obtained {\sl via} the self-consistency
equation~\cite{Cuevas}:
$$
\Delta = U \int_{- \infty}^{+ \infty}
d \omega / (2 i \pi) [ \hat{G}^{+ -}(\omega)]_{2,1}
,
$$
with $U$ the microscopic
attractive interaction. The dominant contribution arises from
the large-$|\omega|$ behavior, which leads to
the BCS-type relation:
\begin{equation}
\label{eq:gap1}
\Delta = D \exp{ \left[ -  \frac{1}{\rho_N U}
\left( 1 + \pi \rho_N \Gamma_\uparrow \right)
\left( 1 + \pi \rho_N \Gamma_\downarrow \right)
\right]}
,
\end{equation}
with $D$ the bandwidth of the superconductor.
As an example, I consider a coupling to two ferromagnets.
With a parallel alignment of the magnetizations in the
electrodes, one has $\Gamma_\uparrow = 2 \gamma$
and $\Gamma_\downarrow = 0$. With an antiparallel
alignment, one has $\Gamma_\uparrow = \Gamma_\downarrow
= \gamma$. The ratio of the two gaps is found to be
\begin{equation}
\label{eq:gap2}
\frac{\Delta_{AP}}{\Delta_P} = \exp{ \left( -
\frac{\pi^2 \rho_N \gamma^2}
{ U } \right)} < 1
,
\end{equation}
which shows that the spin polarized
environment generates
a reduction of the superconducting gap, stronger
with an antiferromagnetic alignment than with
a ferromagnetic
alignment. As a consequence, the transition temperature
of the superconductor is larger if the electrodes
are in a parallel alignment.
This is valid whatever the transparency of the contacts.

To make contact with possible experiments, one should
keep in mind the following points:
\begin{itemize}
\item[(i)] Eq.~\ref{eq:gap2} should be strictly speaking
applied only if the dimension of the superconductor
is small compared to the coherence length.
\item[(ii)] It is nevertheless possible to extrapolate to
the physics occurring when a bulk superconductor is connected
to ferromagnetic electrodes. In this case there is
a non homogeneous gap in the superconductor, which is
reduced at the interfaces with ferromagnetic electrodes.
If two ferromagnetic electrodes are at a distance smaller
than the coherence length, the {\sl local} gap
should follow {\sl qualitatively} Eq.~\ref{eq:gap2}.
A transport experiment
involving two electrodes at a distance smaller
than the coherence length would be sensitive to
the {\sl local} gap.
\end{itemize}

In the remaining of the article, I focus
on transport properties. While keeping in
mind that the superconducting gap should depend
on the spin orientation of the ferromagnetic electrodes,
I will not determine systematically the
superconducting gap in a self consistent manner.
Instead, I assume that the gap is a fixed quantity, but
I could easily inject {\sl a posteriori} the self
consistent value of the
gap in the transport equations.

\subsection{Non local Andreev reflections}

I use Green's function techniques
to evaluate the current $I_n$ flowing into the ferromagnetic electrode
$\alpha_n$ as a trace over Nambu space:
\begin{equation}
I_n = \int d \omega 
\mbox{ Tr} \left\{ \left [
\Ht_{\alpha_n,x} \HG^{+-}_{x,\alpha_n} (\omega) -
\Ht_{x,\alpha_n} \HG^{+-}_{\alpha_n,x} (\omega)
\right] \right\}
.
\label{eq:current}
\end{equation}
Since it is trivial to restore the
quantum of conductance in the transport formula,
I assume that $e/h=1$ throughout the article.

The Hamiltonian decomposes into two contributions:
$$
\hat{\cal H} = \hat{\cal H}_x + \sum_{k=1}^N
\hat{\cal H}_{\alpha_k} + \hat{\cal W}
,
$$
where $\hat{\cal H}_x$ and $\hat{\cal H}_{\alpha_k}$
are associated to the superconducting and
ferromagnetic electrodes, and $\hat{\cal W}$
is the coupling term:
$$
\hat{\cal W} = \sum_{k=1}^N t_{\alpha_k,x}
\left[ c_{\alpha_k}^+ c_x + 
c_x^+ c_{\alpha_k} \right]
.
$$
It is a standard method
to treat $\hat{\cal W}$ as if it were a perturbation,
and to sum to all orders the resulting perturbation theory.
The final transport formula is then {\sl non perturbative}
in $\hat{\cal W}$~\cite{Caroli}. Therefore the method
is well suited to describe
the physics of high transparency contacts.

The chain of Dyson-Keldysh equations associated to
Fig.~\ref{fig:single} takes the form
\begin{eqnarray}
\Ht_{x,\alpha_n} \HG^{+-}_{\alpha_n,x}
&=& \Ht_{x,\alpha_n} \sum_{k=1}^N \HG^R_{\alpha_n,\alpha_k}
\Ht_{\alpha_k,x} \Hg^{+-}_{x,x}
\left[ \HI + \sum_{l=1}^N \Ht_{x,\alpha_l}
\HG^A_{\alpha_l,x} \right]\\
&+& \Ht_{x,\alpha_n} \left[ \HI
+ \HG^R_{\alpha_n,x} \Ht_{x,\alpha_n} \right]
\Hg^{+-}_{\alpha_n,\alpha_n}
\Ht_{\alpha_n,x} \HG^A_{x,x} \\
&+& \sum_{k \ne n}
\Ht_{x,\alpha_n} \HG^R_{\alpha_n,x}
\Ht_{x,\alpha_k} \Hg^{+-}_{\alpha_k,\alpha_k}
\Ht_{\alpha_k,x}
\HG^A_{x,x}
.
\end{eqnarray}
The spin--$\sigma$ Andreev current flowing into
electrode $\alpha_n$ is found to be
\begin{equation}
\label{eq:courant-A}
I^A_{n,\sigma} = 4 \pi^2 
 \Gamma_{n,\sigma} \Gamma_{-\sigma}
\int d \omega \left[n_F(\omega-eV) - n_F(\omega) \right]
|G_{x,x,1,2}^A|^2
.
\end{equation}
The Andreev current Eq.~\ref{eq:courant-A}
is 
similar to the one of conventional Andreev reflection~\cite{Cuevas},
except for the density of state
prefactors appearing in the spectral line-widths.
These prefactors reflect the fact that a spin--$\sigma$
electron can be transfered in electrode $\alpha_n$
only if it paired with a spin--($-\sigma$) electron
either in the same electrode (in which case the
Cooper pair is a local object), or in another
electrode (in which case a non local Cooper
pair is transfered). This shows 
that the Andreev current is built up from
a sum of all possible non local
Cooper pair contributions, and generalizes the situation
considered in Ref.~\cite{Feinberg}.

\section{Effective single site Green's
functions formalism}
\label{sec:forma}
Now I describe the method
that will be used throughout the remaining
of the article. The strategy is to use
Green's functions techniques in an effective single site
formalism. This method has proved to be extremely powerful
in the context of superconducting quantum
point contacts~\cite{Cuevas}. The advantage 
is that simple algebraic expressions can
be manipulated without using spectral representations.
The basic assumption 
is to consider that a given site
represents a phase coherent continuum
of states. 
I first take 
the example of the metal~-- metal~--
metal junction and show how to interpret 
the resulting transport theory.
Since I consider in this section circuits without
superconducting elements, the Nambu space structure
is diagonal, and I can use spinless
fermions without any loss of generality.

\begin{figure}
\centerline{\psfig{file=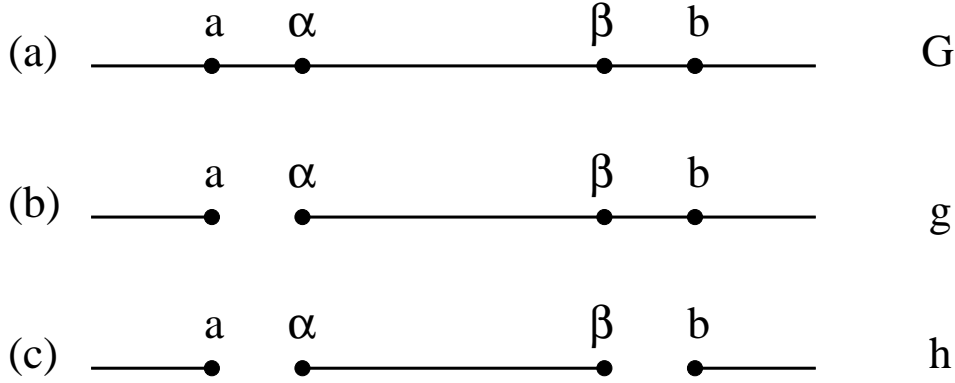,height=5cm}}
\caption{Notations for the Green's functions.
In (a): no link is disconnected;
the corresponding Green's functions
are noted $G_{i,j}$. In (b): the link
$a$~--~$\alpha$ is disconnected;
the corresponding
Green's functions are noted $g_{i,j}$.
In (c): The two links $a$~--~$\alpha$ and $b$~--~$\beta$ 
are disconnected. The Green's functions are noted $h_{i,j}$.
One has $g_{a,a}=
h_{a,a}$.
\base
}
\label{fig:app}
\end{figure}

\subsection{Treating all couplings on the same footing}
\label{sec:1step}
Let us start to calculate the transport formula
corresponding to the metal~-- metal~-- metal
junction shown on Fig.~\ref{fig:app}. 
We note $G$ the Green's function of the fully
connected system, $g$ the Green's function with
$t_{a,\alpha}=0$, and $h$ the Green's function
with $t_{a,\alpha}=t_{b,\beta}=0$
(see Fig.~\ref{fig:app}).

I treat simultaneously the two couplings
$t_{a,\alpha}$ and $t_{b,\beta}$ on the same
footing. Namely, I go directly
from the $G$'s to the $h$'s without  ever using
the $g$'s.
The chain of Dyson equations for the Green's
function takes the form
\begin{eqnarray}
\label{eq:G-dir1}
G_{a,a} &=& h_{a,a} + h_{a,a} t_{a,\alpha}
h_{\alpha,\alpha} t_{\alpha,a} G_{a,a} + 
h_{a,a} t_{a,\alpha}
h_{\alpha,\beta} t_{\beta,b}
G_{b,a} \\
G_{b,a} &=& h_{b,b} t_{b,\beta}
h_{\beta,\beta} t_{\beta,b} G_{b,a} +
h_{b,b} t_{b,\beta}
h_{\beta,\alpha} t_{\alpha,a} G_{a,a}
\label{eq:G-dir4}
.
\end{eqnarray}
The effective single site Green's functions
are
$h_{a,a}^{A,R} = \pm i \pi \rho_a$,
$h_{i,j}^{A,R} = \pm i \pi \rho$
($i,j=\alpha,\beta$),
$h^{A,R}_{b,b} = \pm i \pi \rho_b$.
I use the notation
$\gamma_a=\pi^2 |t_{a,\alpha}|^2
\rho_a$ and
$\gamma_b=\pi^2 |t_{b,\beta}|^2
\rho_b$. Eqs.~\ref{eq:G-dir1}~--
\ref{eq:G-dir4} are solved in a
straightforward fashion. For instance, one
obtains
$$
G_{a,a}^{A,R} = \pm i \pi \rho_a
\frac{1 + \rho \gamma_b}
{1 + \rho ( \gamma_a + \gamma_b)}
.
$$
Next, I calculate the Keldysh component
\begin{eqnarray}
G_{a,\alpha}^{+-} &=& \left[ 1 + G_{a,\alpha}^R
T_{\alpha,a} \right] h_{a,a}^{+-}
t_{a,\alpha} G_{\alpha,\alpha}^A
+ G_{a,a}^R t_{a,\alpha} h_{\alpha,\alpha}^{+-}
\left[ 1 + t_{\alpha,a} G_{a,\alpha}^A \right]
+ G_{a,a}^R t_{a,\alpha} h_{\alpha,\beta}^{+-}
t_{\beta,b} G_{b,\alpha}^A \\
&+& G_{a,b}^R t_{b,\beta} h_{\beta,\alpha}^{+-}
\left[ 1 + t_{\alpha,a} G_{a,\alpha}^A \right]
+ G_{a,b}^R t_{b,\beta} h_{\beta,\beta}^{+-}
t_{\beta,b} G_{b,\alpha}^A
+ G_{a,\beta}^R t_{\beta,b} h_{b,b}^{+-}
t_{b,\beta} G_{\beta,\alpha}^A
.
\end{eqnarray}
It is straightforward to evaluate
the six terms to obtain the transport formula
\begin{equation}
\label{eq:tr-spinless}
I_{a,\alpha} = \int d \omega \frac{2 \rho \gamma_a
(1 + \rho \gamma_b)}
{ \left[ 1 + \rho (\gamma_a
+ \gamma_b) \right]^2}
\left[ n_F(\omega - eV)
- n_F(\omega) \right]
.
\end{equation}

\subsection{Decimating part of the circuit}
\label{sec:2step}
Before discussing the physics of Eq.~\ref{eq:tr-spinless},
I give another derivation of this
transport formula.
The idea is to use first a perturbation
in $t_{b,\beta}$ and, in a second step, make
a perturbation in $t_{a,\alpha}$.
Namely, part of the circuit is decimated exactly,
and I look for an effective theory for the
remaining sites.

The perturbation in $t_{a,\alpha}$ leads to
\begin{equation}
\label{eq:G-truc}
G^{+-}_{a,\alpha} =
G_{a,a}^R t_{a,\alpha} g_{\alpha,\alpha}^{+-}
\left[ 1 + t_{\alpha,a} G_{a,\alpha}^A \right]
+ \left[ 1 + G_{a,\alpha}^R t_{\alpha,a} \right]
g_{a,a}^{+-} t_{a,\alpha}
G_{\alpha,\alpha}^A
.
\end{equation}
Therefore, I need to determine
$g_{\alpha,\alpha}^{+-}$ and
$g_{a,a}^{+-}$ in terms of the $h$'s.
The chain of Dyson equations reads
\begin{eqnarray}
\label{eq:g-truc1}
g_{\alpha,\alpha} &=& h_{\alpha,\alpha} + h_{\alpha,\beta}
t_{\beta,b} g_{b,\alpha}\\
\label{eq:g-truc2}
g_{b,\alpha} &=& h_{b,b} t_{b,\beta}
g_{\beta,\alpha} \\
\label{eq:g-truc3}
g_{\beta,\alpha} &=& h_{\beta,\alpha} +
h_{\beta,\beta} t_{\beta,b} g_{b,\alpha}
.
\end{eqnarray}
The equation for the Keldysh component is
\begin{equation}
\label{eq:gpm-truc}
g_{\alpha,\alpha}^{+-} = g_{\alpha,b}^R t_{b,\beta} h_{\beta,\beta}^{+-}
t_{\beta,b} g_{b,\alpha}^A
+ h_{\alpha,\alpha}^{+-}
+ h_{\alpha,\beta}^{+-} t_{\beta,b} g_{b,\alpha}^A
+ g_{\alpha,b}^R t_{b,\beta} h_{\beta,\alpha}^{+-}
+ g_{\alpha,\beta}^R t_{\beta,b}
h_{b,b}^{+-} t_{b,\beta}
g_{\beta,\alpha}^A
.
\end{equation}
It is straightforward to solve
Eqs.~\ref{eq:g-truc1}~-- \ref{eq:gpm-truc},
inject the propagators into Eq.~\ref{eq:G-truc}
and find the current Eq.~\ref{eq:tr-spinless}.
In this approach, the effect of the coupling
$\beta$--b is just to
renormalize the density of states at site
$\alpha$:
$g^{+ -}_{\alpha,\alpha} = 2 i \pi n_F(\omega)
\tilde{\rho}_\alpha$,
$g^{A,R}_{\alpha,\alpha} = \pm i \pi
\tilde{\rho}_\alpha$, with
\begin{equation}
\tilde{\rho}_\alpha =\frac{\rho}{1 + \rho \gamma_b}
,
\label{eq:CG}
\end{equation}
where $\rho$ is the density of states
in the intermediate region and $\gamma_b=\pi^2
|t_{b,\beta}|^2 \rho_b$.
We will apply this decimation method latter
in more complicated situations.

\begin{figure}
\centerline{\psfig{file=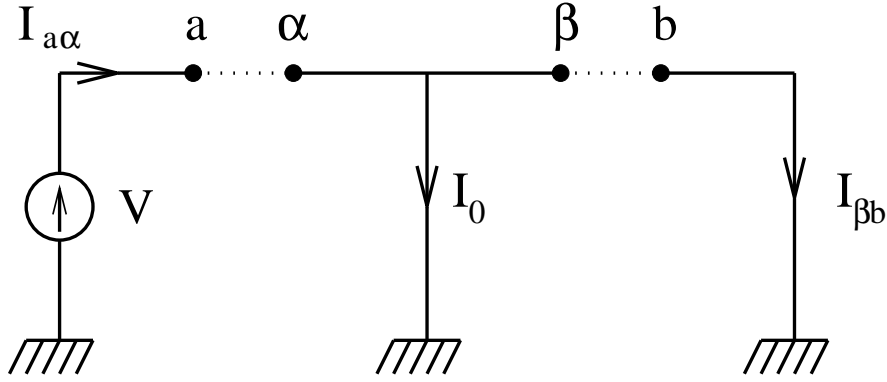,height=5cm}}
\caption{Representation of the current flow.
$I_0$ is a disconnected current and only
$I_{\beta,b}$ is relevant.
\base
}
\label{fig:drain}
\end{figure}
\begin{figure}
\centerline{\psfig{file=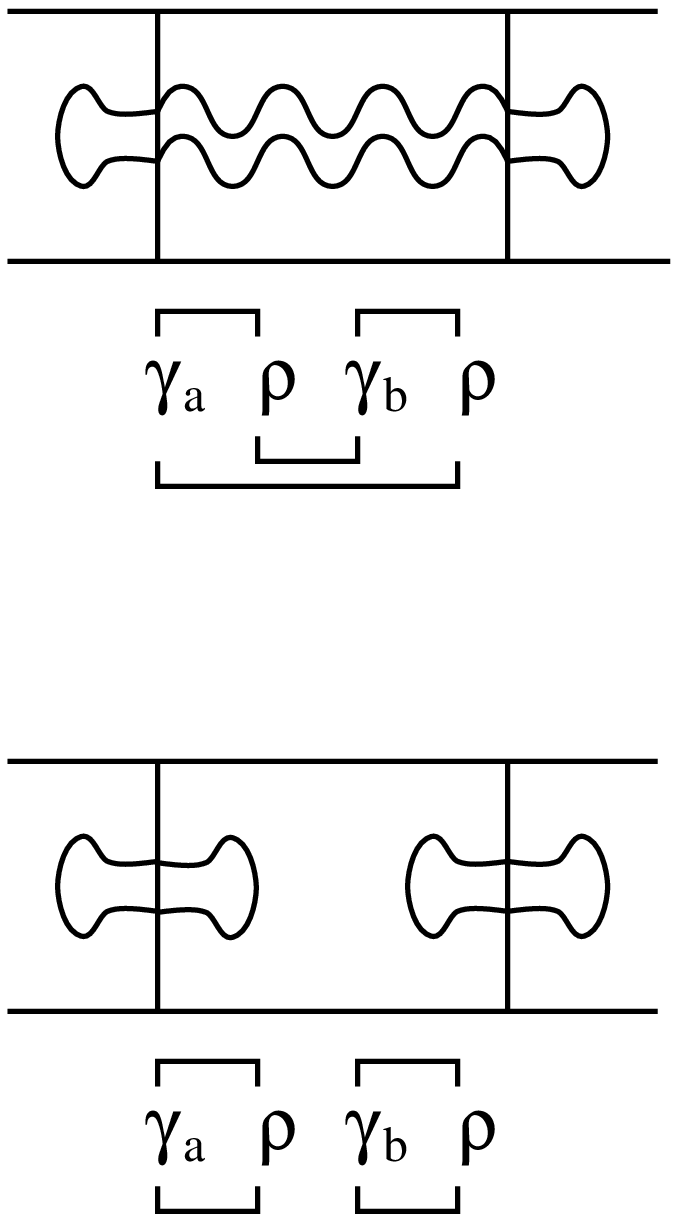,height=9cm}}
\caption{The diagrams of order $\rho^2
\gamma_a \gamma_b$
contributing to $I_{\beta,b}$. The top
diagram  has
a unit charge ($Q=1$). The bottom
diagram has a zero charge ($Q=0$).
\base
}
\label{fig:order2}
\end{figure}
\subsection{Interpretation of the transport formula}
The current Eq.~\ref{eq:tr-spinless} can be decomposed
into two contributions: $I_{a,\alpha} = I_0 + I_{\beta,b}$, with
\begin{equation}
\label{eq:I0}
I_0 = \int d \omega \frac{2 \rho \gamma_a}
{ \left[ 1 + \rho ( \gamma_a + \gamma_b) \right]^2}
\left[ n_F(\omega-eV) - n_F(\omega) \right]
,
\end{equation}
and
\begin{equation}
\label{eq:Ibetab}
I_{\beta,b} = \int d \omega \frac{2 \rho^2 \gamma_a \gamma_b}
{ \left[ 1 + \rho ( \gamma_a + \gamma_b) \right]^2}
\left[ n_F(\omega-eV) - n_F(\omega) \right]
.
\end{equation}
Is is straightforward to calculate directly the
current flowing through the link $\beta$~-- $b$,
and show that it coincides with Eq.~\ref{eq:Ibetab}.
Therefore, we are lead to consider
that the region $\alpha$~--~$\beta$ can absorb
part of the electrical current
(see Fig.~\ref{fig:drain}). Only is the
transmitted current $I_{\beta,b}$ a quantity
of interest. To understand the appearance
of the fake term $I_0$, it is
useful to notice that
an effective single site plays two
roles at the same time:
\begin{itemize}
\item[(i)] It transmits current to the rest
of the circuit. This is why there is a finite
contribution $I_{\beta,b}$.
\item[(ii)] It represents a continuum of
energy levels; therefore, it also plays
by itself the role of a reservoir. This is
why there is a finite current $I_0$.
\end{itemize}

Now the expression (\ref{eq:Ibetab}) of $I_{\beta,b}$
contains also a renormalization prefactor which
generates an infinite series of Feynman diagrams.
It turns out that disconnected contributions
are also present in these diagrams.
To obtain the final form of the physical current,
we expand 
Eq.~\ref{eq:Ibetab} in an infinite series
of Feynman diagrams, and operate on this
series with an operator $\hat{Q}$ that
multiplies each Feynman diagram its charge:
\begin{equation}
\label{eq:Iphys1}
I_{\rm phys} = \int d \omega \hat{Q}
\left[ \frac{2 \rho^2 \gamma_a \gamma_b}
{ \left[ 1 + \rho ( \gamma_a + \gamma_b) \right]^2}
\right]
\left[ n_F(\omega-eV) - n_F(\omega) \right]
.
\end{equation}

In the case of the metal~--metal~--metal junction,
Eq.~\ref{eq:Iphys1} should coincide
with the transport formula obtained by Caroli
{\sl et al.}~\cite{Caroli}:
\begin{equation}
\label{eq:Caroli}
I_{\rm phys} = \int d \omega 
\frac{\rho^2 \gamma_a \gamma_b}
{ \left[ 1 + \rho ( \gamma_a + \gamma_b) \right]^2}
\left[ n_F(\omega-eV) - n_F(\omega) \right]
.
\end{equation}
To verify this, I calculate the lowest
order diagrams associated to Eq.~\ref{eq:Iphys1}.
This will give us the opportunity to show
how the operator $\hat{Q}$ operates.
\begin{figure}
\centerline{\psfig{file=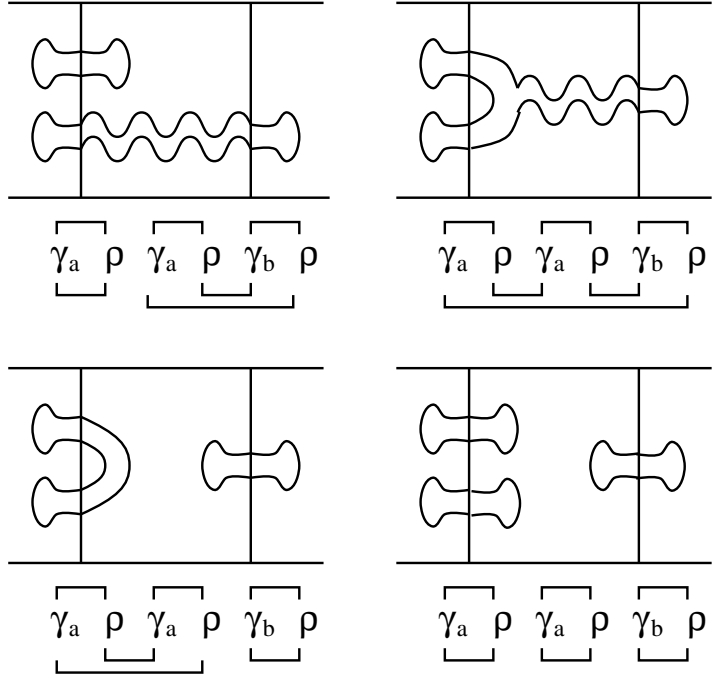,height=9cm}}
\caption{The four diagram of order $\rho^3
\gamma_a^2 \gamma_b$
contributing to $I_{\beta,b}$. The two top
diagrams has a unit charge ($Q=1$). 
The two bottom diagrams have a zero charge ($Q=0$).
\base
}
\label{fig:order3}
\end{figure}

\subsection{Feynman diagrams}

\begin{figure}
\centerline{\psfig{file=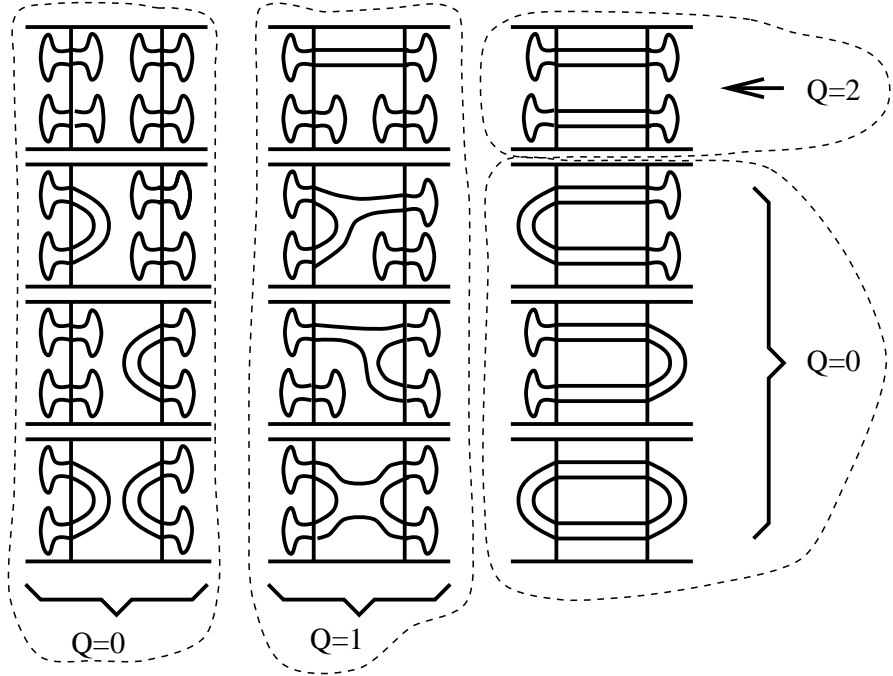,height=9cm}}
\caption{The twelve diagram of order $\rho^4
\gamma_a^2 \gamma_b^2$
contributing to $I_{\beta,b}$. The charges 
are indicated.
\base
}
\label{fig:12}
\end{figure}

To count how many diagrams appear
at a given order,
we expand Eq.~\ref{eq:Ibetab} in a power series:
\begin{equation}
\label{eq:expan1}
\frac{2 \rho^2 \gamma_a \gamma_b}
{ \left( 1 + \rho(\gamma_a + \gamma_b)
\right)^2} = 
2 \rho^2 \gamma_a \gamma_b - 4 \rho^3 \gamma_a
\gamma_b \left( \gamma_a + \gamma_b\right)
+ 6 \rho^4 \gamma_a \gamma_b
\left( \gamma_a + \gamma_b \right)^2
+ ...
.
\end{equation}
From what we deduce that there are two
diagrams at order $\gamma_a \gamma_b$,
four diagrams at order $\gamma_a^2 \gamma_b$,
six diagrams
at order $\gamma_a^3 \gamma_b$, twelve
diagrams at order $\gamma_a^2 \gamma_b^2$,
etc ...
Each of the diagrams corresponds to a possible
contraction of the density of states.
For instance the two diagrams at order
$\gamma_a \gamma_b$ are shown on Fig.~\ref{fig:order2}.
The four diagrams at order $\gamma_a^2 \gamma_b$
are shown on Fig.~\ref{fig:order3}.
The twelve diagrams at order $\gamma_a^2 \gamma_b^2$
are shown on Fig.~\ref{fig:12}.
Using these diagrams to evaluate Eq.~\ref{eq:Iphys1}, we obtain
the expansion
$$
I_{\rm phys} = \int d \omega
\left[ \rho^2 \gamma_a \gamma_b
-2 \rho^3 \gamma_a \gamma_b ( \gamma_a
+ \gamma_b) + 3 \rho^4
\gamma_a \gamma_b ( \gamma_a + \gamma_b)^3
+ ...
\right] \times
\left[ n_F(\omega-eV) - n_F(\omega) \right]
,
$$
which coincides with the expansion of
Eq.~\ref{eq:Caroli}.

\section{Transport of linear superposition
of correlated pairs}
\label{sec:ent}

\subsection{EPR paradox}
It is worth recalling 
the well known EPR paradox~\cite{EPR}.
Let us consider a collision of two spin--$1/2$ electrons.
The collision is
represented by an antiferromagnetic
exchange interaction $J(\tau) {\bf S}_1 . {\bf S}_2$,
where $J(\tau)$ is constant in the time interval
$\left[ 0, t \right]$ and zero otherwise.
After the collision, the system is represented
by the EPR state
\begin{equation}
\label{eq:psi-EPR}
| \psi \rangle = \cos{(\Omega t)} | + - \rangle
- i \sin{(\Omega t)} | - + \rangle
,
\end{equation}
with $\Omega=J/2$. Let us imagine that two observers
${\cal O}_1$ and ${\cal O}_2$ use an apparatus
({\sl i.e.} a Stern and Gerlach magnet)
to measure the spin orientation of ${\bf S}_1$
and ${\bf S}_2$.
The observator ${\cal O}_1$ measuring
$S_1^z$ would find $1/2$ with a probability
$\cos^2{(\Omega t)}$ and $-1/2$ with a probability
$\sin^2{(\Omega t)}$.
Once the observer ${\cal O}_1$ has made his measurement,
the wave function collapses and with a probability one,
the observer ${\cal O}_2$
measuring $S_2^z$ would find
$S_2^z=1/2$ if ${\cal O}_1$ has measured $S_1^z=-1/2$,
and $S_2^z=-1/2$ if ${\cal O}_1$ has measured $S_1^z=1/2$.
Quantum mechanics contains non separable correlations,
which is the famous EPR paradox~\cite{EPR}. As pointed
out by Bell, there exists a basic difference between
quantum mechanics and hidden variable theories~\cite{Bell}.
It has been
demonstrated experimentally that Bell inequalities
were violated with photons~\cite{Aspect,Kwiat}, therefore
ruling out hidden variable theories (at least with photons).
As already
mentioned in the Introduction, it is an open question
to test the EPR paradox with electrons. We do not
address this issue directly
in this article. Nevertheless,
we find an interesting connection between transport
properties and quantum measurement ({\sl i.e.}
projections) of linear superpositions of correlated pairs.

\begin{figure}
\centerline{\psfig{file=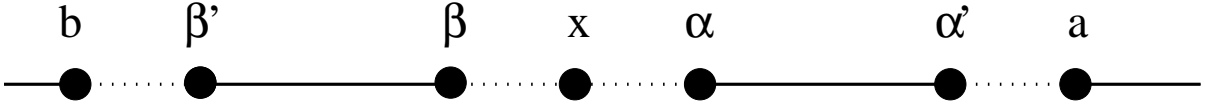,height=1.3cm}}
\caption{Schematic representation of the 
model considered in section~\ref{sec:ent}. The site
$x$ is superconducting. The intermediate regions
$\alpha$~--~$\alpha'$ and $\beta$~--~$\beta'$
are ferromagnetic.
The electrodes $a$ and $b$ are ferromagnetic.
\base
}
\label{fig:double}
\end{figure}

\subsection{Fabrication of linear superpositions
of correlated pairs}
\label{sec:fab}
\begin{figure}
\centerline{\psfig{file=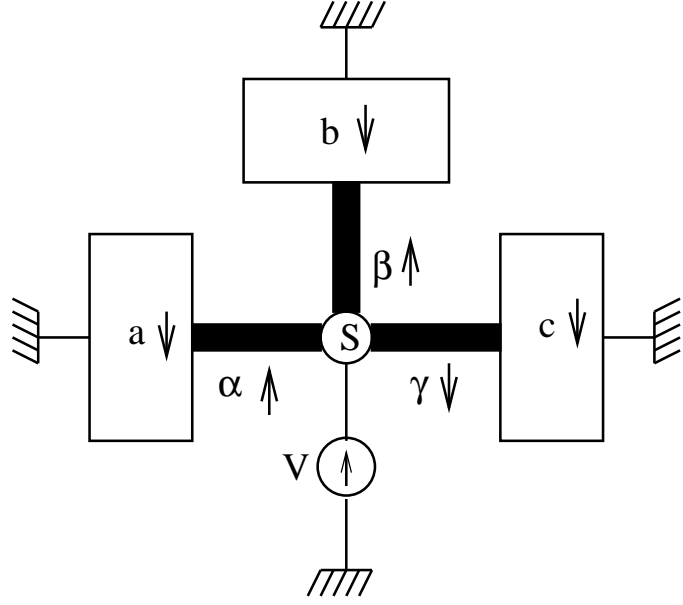,height=8cm}}
\caption{Representation of a device in which three
ferromagnetic electrodes $\alpha$, $\beta$, $\gamma$
are connected to a superconductor. The additional
electrodes a, b, c are used to
to perform a measurement of the linear superposition
of correlated pairs.
\base
}
\label{fig:3ferro}
\end{figure}

Now I use the formalism presented in section~\ref{sec:forma}
to discuss the physics of linear superpositions
of correlated pairs of electrons. I consider that a
ballistic ferromagnetic region $\alpha_k$~--~$\alpha_k'$
is inserted in
between the superconducting site $x$ and the external
ferromagnetic electrodes $a_k$ (see Figs.~\ref{fig:double}
and~\ref{fig:3ferro}).

To simplify the discussion, I make the following assumptions:
\begin{itemize}
\item[(i)]  There are no pair correlations
between the bottom electrode on Fig.~\ref{fig:double}
(used to impose the superconductor chemical potential)
and the other ferromagnetic electrodes.
This means that the corresponding contact
on the superconductor
is separated from the other contacts by a distance
much larger than the superconducting coherence length.

\item[(ii)] All ferromagnets are fully polarized.
Therefore, there is no transmission of local
Cooper pairs.

\item[(iii)]
The ferromagnetic intermediate regions are phase
coherent. Present time technology
does not allow the fabrication of such devices. Nevertheless,
the model is interesting {\sl per se}.
\end{itemize}

Our analysis of the situations presented on
Figs.~\ref{fig:double} and~\ref{fig:3ferro} is
the following.
The wave function in the ballistic regions is
a linear superposition of correlated pairs.
For instance, in the presence of
three fully polarized
ferromagnetic regions connected to the superconductor
having a spin orientation $\sigma_\alpha=\uparrow$,
$\sigma_\beta=\sigma_\gamma=\downarrow$ (see Fig.~\ref{fig:3ferro}),
the wave function is
\begin{equation}
\label{eq:psi}
| \psi \rangle =
c_{\alpha, \uparrow}^+ \left[
\sqrt{ \frac{t_\beta \rho_{\beta,\downarrow}}
{t_\beta \rho_{\beta,\downarrow} +
t_\gamma \rho_{\gamma,\downarrow}}}
c_{\beta,\downarrow}^+
+
\sqrt{ \frac{t_\gamma \rho_{\gamma,\downarrow}}
{t_\beta \rho_{\beta,\downarrow}+ t_\gamma
\rho_{\gamma,\downarrow}}}
c_{\gamma,\downarrow}^+
\right] |0 \rangle
.
\end{equation}
In the presence of spin orientations $\sigma_\alpha=
\sigma_\beta=\uparrow$, $\sigma_\gamma=\downarrow$,
we have
\begin{equation}
\label{eq:psi2}
| \psi' \rangle =
\left[
\sqrt{ \frac{t_\beta \rho_{\beta,\uparrow}}
{t_\alpha \rho_{\alpha,\uparrow}+
t_\beta \rho_{\beta,\uparrow}}}
c_{\beta,\uparrow}^+
+
\sqrt{ \frac{t_\alpha \rho_{\alpha,\uparrow}}
{t_\alpha \rho_{\alpha,\uparrow}+
t_\beta \rho_{\beta,\uparrow}}}
c_{\alpha,\uparrow}^+
\right] c_{\gamma,\downarrow}^+
|0 \rangle
.
\end{equation}
These wave functions are the only ones that can
guarantee the correct pairing associated
to the formation of correlated pairs of electrons.
The coefficients of the wave function can be obtained by
evaluating
the Gorkov function. With the configuration
$\sigma_\alpha=\sigma_\beta=\uparrow$,
$\sigma_\gamma=\downarrow$, we find
$$
\left[ \hat{G}_{\alpha (\beta),\gamma}^{+ -} \right]_{1,2}
= \pi^2 t_{\alpha (\beta)} t_\gamma
\rho_{\alpha (\beta),\uparrow}
\rho_{\gamma,\downarrow}
\left[ \hat{G}_{x,x}^{+ -} \right]_{1,2}
.
$$ 
From what we deduce Eq.~\ref{eq:psi}.
We will come back in section~\ref{sec:measurement}
on the meaning of this wave function.

The dictionary between
the Cooper pair wave function Eq.~\ref{eq:psi}
and the spin--$1/2$ wave function Eq.~\ref{eq:psi-EPR}
is the following:
\begin{eqnarray}
|+ - \rangle &\leftrightarrow& 
c_{\alpha, \uparrow}^+  c_{\beta,\downarrow}^+\\
|- + \rangle &\leftrightarrow& 
c_{\alpha, \uparrow}^+  c_{\gamma,\downarrow}^+
.
\end{eqnarray}
The relevant question that should be asked now is
to determine whether it is possible to make a
quantum measurement of the linear superposition of
correlated pairs, similar to the Stern and
Gerlach measurement of the spin-$1/2$
EPR state. Namely, we want to interpret transport
theory in terms of projection operators.

\subsection{Transport formula}
\label{sec:calcul}

\begin{figure}
\centerline{\psfig{file=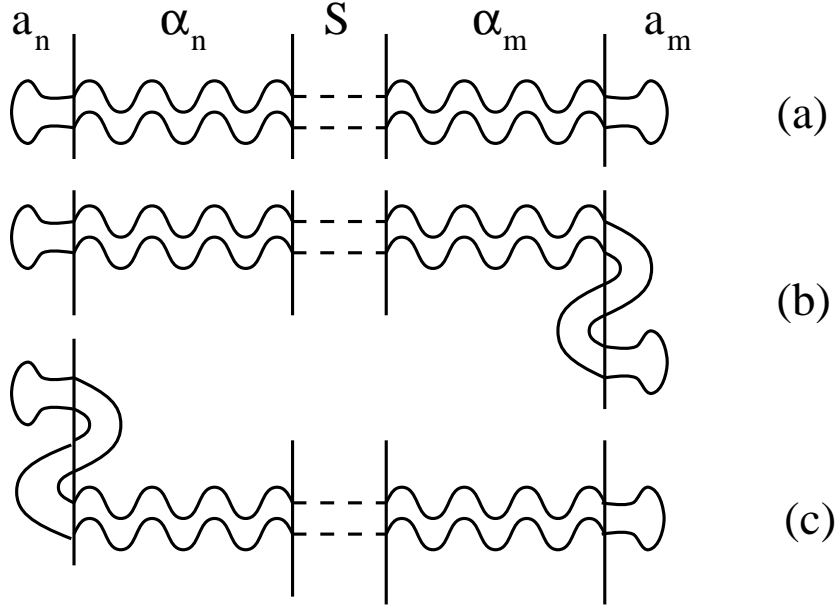,height=8cm}}
\caption{The relevant lowest order Andreev reflection diagrams.
(a) is of order $\rho_{a_n} \rho_{a_m} (\rho_{\alpha_n})^2
(\rho_{\alpha_m})^2$.
(b) is of order $\rho_{a_n} (\rho_{a_m})^3 (\rho_{\alpha_n})^2
(\rho_{\alpha_m})^4$.
(c) is of order $(\rho_{a_n})^3 \rho_{a_m} (\rho_{\alpha_n})^4
(\rho_{\alpha_m})^2$. The dashed line represents
the anomalous propagation in the superconductor.
\base
}
\label{fig:A-relevant}
\end{figure}

The strategy to solve the model
on Fig.~\ref{fig:double} within the effective
single site Green's function approach is
to use the two step perturbation
theory presented in section~\ref{sec:2step}.
I first treat the link
$\alpha'$~--~$a$ in perturbation (see Fig.~\ref{fig:double}).
Since the superconducting site has been disconnected
and there is no source of spin flip scattering,
the Dyson equations decouple into a spin-up
and a spin-down component.
We note $h_{i,j}$ the Green's functions
of the disconnected system, {\sl i.e.} with
$t_{a,\alpha_k}=0$. 
The Dyson equation for the Keldysh component
has already been given in section~\ref{sec:forma}.
The solution takes the form
$g_{\alpha,\alpha,\sigma}^{+-} = 2 i \pi n_F(\omega)
\tilde{\rho}_{\alpha,\sigma}$, and
$g_{\alpha,\alpha,\sigma}^{A,R} = \pm i \pi
\tilde{\rho}_{\alpha,\sigma}$
,
with the renormalized density of states
$$
\tilde{\rho}_{\alpha,\sigma} = \frac{ 
\rho_{\alpha,\sigma} }
{ 1 + \pi^2 |t_{a,\alpha}|^2 \rho_{\alpha,\sigma}
\rho_{a,\sigma} }
.
$$
Next, I use a perturbation in $t_{x,\alpha_k}$
to calculate the Andreev current. This is done
by noticing that the physical
current $I_{\alpha,a}$ is a finite fraction
of the current incoming at site $\alpha$
(see Eqs.~\ref{eq:I0},~\ref{eq:Ibetab}):
\begin{equation}
\label{eq:ratio}
\frac{ I_{\alpha,a}}{I_{x,\alpha}}
= \frac{\pi^2 |t_{a,\alpha}|^2 \rho_a \rho_\alpha}
{ 1 + \pi^2 |t_{a,\alpha}|^2 \rho_a \rho_\alpha}
.
\end{equation}
I deduce from Eq.~\ref{eq:ratio} that 
the Andreev current Eq.~\ref{eq:courant-A} is the sum
of all possible Cooper pair transmissions:
\begin{equation}
\label{eq:Landauer}
I^A_{n,\sigma} = 4 \pi^2 
\int d \omega \left[ n_F(\omega-eV)
- n_F(\omega) \right] \sum_{m}  
\hat{Q} \left[
u_{n,\sigma,\sigma'}
u_{m,-\sigma,\sigma'}\right]
|G_{x,x,1,2}^A|^2
,
\end{equation}
where
\begin{equation}
\label{eq:u}
u_{n,\sigma,\sigma'} = \pi^2
|t_{x,\alpha_n}|^2 |t_{a_n,\alpha_n}|^2
\frac{\rho_{\alpha_n,\sigma}^2 \rho_{a_n,\sigma}}
{ \left( 1 + \pi^2 |t_{a_n,\alpha'_n}|^2
\rho_{\alpha_n,\sigma}
\rho_{a_n,\sigma}  \right)^2}
.
\end{equation}
The action of $\hat{Q}$ in Eq.~\ref{eq:Landauer}
has already been given.
The form Eq.~\ref{eq:Landauer} 
of the current can also be obtained by
making directly a perturbation in $t_{x,\alpha_k}$
to evaluate the current through the link
$\alpha_n$~--~$a_n$. The
density of state prefactors appearing in
the lowest contribution in Eq.~\ref{eq:Landauer}
corresponds to the diagram on
Fig.~\ref{fig:A-relevant}~(a).

\subsection{Quantum measurement of the superposition
of correlated pairs}
\label{sec:measurement}

Let us now interpret the
transport formula Eq.~\ref{eq:Landauer}
in terms of quantum measurement of the superposition
of correlated pairs.
We assume that the 
external ferromagnetic electrodes are
weakly coupled to the intermediate regions.
As a consequence, we make the approximation
$\tilde{\rho}_\alpha = \rho_\alpha$.
The interfaces with the superconductor can
be either tunnel or high transparency contacts.
The general form of the linear superposition
is
\begin{equation}
\label{eq:psi-general}
| \psi \rangle =
{\cal N}^{-1/2}
\sum_{p,q} \sqrt{ t_{\alpha_p}
t_{\alpha_q} \rho_{\alpha_p,\uparrow}
\rho_{\alpha_q,\downarrow}}
c_{\alpha_p,\uparrow}^+
c_{\alpha_q,\downarrow}^+
| 0 \rangle
,
\end{equation}
where
$$
{\cal N} = \sum_{p} t_{\alpha_p}
\rho_{\alpha_p,\uparrow} 
\sum_{q} t_{\alpha_q}
\rho_{\alpha_q,\downarrow}
.
$$
It is useful to define
a projection operator associated to
the correlated pairs of electrons $(p \uparrow,q
\downarrow)$:
$$
\hat{\cal P}_{p,q}
= c_{\alpha_{p},\uparrow}^+
c_{\alpha_{q},\downarrow}^+
c_{\alpha_{p},\uparrow}
c_{\alpha_{q},\downarrow}
.
$$
The spin-up current through electrode
$a_p$ can be rewritten in the form
\begin{equation}
\label{eq:current-proj}
I_p= 4 \pi^2 {\cal N}^2
\sum_{q}
\int d \omega \left[ n_F(\omega-eV)
-n_F(\omega) \right] \gamma_{a_p,\uparrow}
\gamma_{a_q,\downarrow}
\left| \langle \psi | 
\hat{\cal P}_{p,q}
| \psi \rangle \right|^2
|G_{x,x,1,2}^A|^2
,
\end{equation}
where $\gamma_{a_{p (q)},\uparrow} = 
\pi^2 |t_{a_{p (q)}}|^2 \rho_{a_{p (q)},\uparrow
(\downarrow)}$
denotes the spectral line-width associated
to the external electrodes.
We first notice that the current is a sum
over all possible Cooper pair transmissions,
which we already found in
section~\ref{sec:1}. What is new here is that
there is  one intermediate quantum state
is involved
(the linear superposition). The transport 
formula can be expressed as suitable
projections of this quantum state. These
projections correspond to the physical processes
by which current is carried away from the
quantum state, namely the transport of
correlated pairs of electrons.
Therefore, the whole picture is consistent:
we first identified a wave function, and next showed that
transport can be interpreted in terms of quantum measurement
of the wave function. In our opinion, the form
(\ref{eq:current-proj}) of the transport
formula is not specific to our model, but is valid
independently on the microscopic origin of
the pairing potential. 

This behavior is also consistent with
general considerations on
decoherence~\cite{Unruh:1989-1,Zurek:1982-1,Zurek:1991-1,degio},
at least on a heuristic basis.
I conjecture that there are
three distinct stages involved in the decoherence mechanism,
which occur on different time scales:
\begin{itemize}
\item[(A)] The initial quantum
superposition Eq.~\ref{eq:psi-general}, being a Schr\"odinger cat
of correlated pairs of electrons.
\item[(B)] An intermediate stage in which the Schr\"odinger cat
superposition is lost but
there are still pair correlations. Namely,
the spin-up (-down) electrons have tunneled into
the specific reservoirs $p$ ($q$). The correlated
pair still behaves like a two-electron phase
coherent object.
\item[(C)] The final stage in which the two electrons
making the correlated pair behave like independent
quasi-particles, with no more pair correlations.
\end{itemize}
The transition (B)~$\rightarrow$ (C) is usually known as
the proximity effect.
I conjecture that the projector
appearing in Eq.~\ref{eq:current-proj} is
the signature of the decoherence process
(A)~$\rightarrow$ (B). Note that because two-electron
correlations are preserved until stage (C),
there is an interesting physics going on
at stage (B) which is explored in the next
section.

The reader might object that the
intermediate regions may contain thousands
of electrons. It may then appear counter-intuitive
to propose Schr\"odinger cat made of
superpositions of a large number of
electrons. This is certainly true, and one should
keep in mind that
{\sl the wave function Eq.~\ref{eq:psi-general}
is by no way the many body ground state
wave function}. Instead it is intended to describe
only the degrees of freedom that participate
to transport.

\begin{figure}
\centerline{\psfig{file=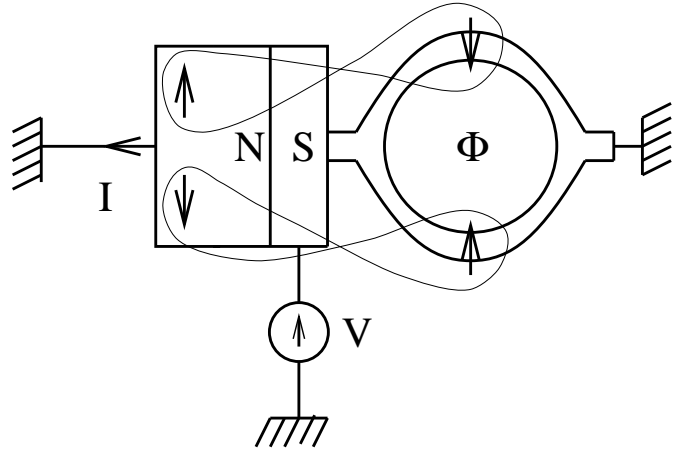,height=6cm}}
\caption{Geometry of the Aharonov-Bohm experiment.
The current $I$ flowing into the left electrode
is modulated by the flux enclosed by the other electron
making the correlated pair. The non separable
correlations between the left electrode and the
Aharonov-Bohm loop are represented schematically.
\base
}
\label{fig:AB}
\end{figure}

\section{Aharonov-Bohm effect}
\label{sec:AB}
Now I propose a device that can be used to 
probe non separable correlations. 
For this purpose, I consider
the geometry on Fig.~\ref{fig:AB},
in which one side of the superconductor is
connected to an Aharonov-Bohm loop (being made
of a normal metal), while the
other side is connected to a normal metal that
does not contain any hole.
I show that the current flowing through the
electrode with no hole is modulated by the flux
enclosed by the normal metal loop. 
This means that the current due to one electron
making a correlated pair is modulated by the flux
enclosed by the other electron, which is a manifestation
of non separable correlations.
The electrodes are considered to be normal metals
in which phase coherence can propagate over large
distances.

\begin{figure}
\centerline{\psfig{file=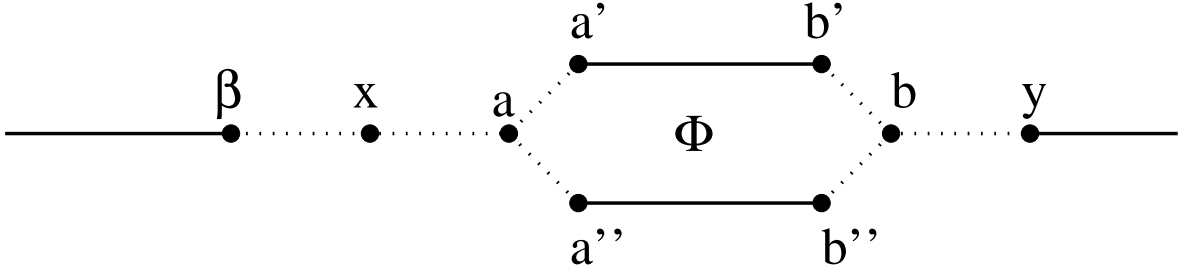,height=3.5cm}}
\caption{The circuit used to represent the
situation on Fig.~\ref{fig:AB}. $x$ is superconducting,
$\beta$ is ferromagnetic and the remaining
sites are normal metals.
\base
}
\label{fig:AB-th}
\end{figure}

I consider the circuit on Fig.~\ref{fig:AB-th}
that is used to model the situation on Fig.~\ref{fig:AB}.
The hopping parameters are complex number, for instance
$t_{a,a'}=t \exp{(i \varphi_{a,a'})}= t
\exp{\left[2 i \pi \phi / (4 \phi_0)\right]}$,
and $t_{a',a} = t \exp{(i \varphi_{a',a})}=
t \exp{\left[-2 i \pi
\phi/(4 \phi_0)\right]}$.
Since I use effective single site Green's functions
in which $a'$ and $a''$ are represented by
the same single site,
this choice of the hopping insures
that $\oint \varphi_{i,j} =  2 \pi 
\phi / \phi_0$.
The first task is to solve the chain of Dyson
equations
\begin{eqnarray}
\label{eq:Dyson-ini}
G_{a,a} &=& g_{a,a} + g_{a,a}
t_{a,a'} G_{a',a} +
g_{a,a} t_{a,a''} G_{a'',a}\\
G_{a',a} &=& g_{a',a'} t_{a',a} G_{a,a}
+ g_{a',b'} t_{b',b} G_{b,a} \\
G_{a'',a} &=& g_{a'',a''} t_{a'',a} G_{a,a}
+ g_{a'',b''} t_{b'',b} G_{b,a} \\
G_{b,a} &=& g_{b,b} t_{b,b'} G_{b',a}
+ g_{b,b} t_{b,b''} G_{b'',a} \\
G_{b',a} &=& g_{b',b'} t_{b',b} G_{b,a} 
+ g_{b',a'} t_{a',a} G_{a,a} \\
G_{b'',a} &=& g_{b'',b''} t_{b'',b}
G_{b,a} + g_{b'',a''} t_{a'',a} G_{a,a}
\label{eq:Dyson-fin}
.
\end{eqnarray}
Next we need to calculate the Keldysh component
\begin{eqnarray}
\label{terme1}
G^{+-}_{a,a} &=& \left[ 1 + G_{a,a}^R t_{a',a}
+ G_{a,a''}^R t_{a'',a} \right]
g_{a,a}^{+-}
\left[ 1 + t_{a,a'} G_{a',a}^A + t_{a,a''}
G_{a'',a}^A \right]\\
\label{terme2}
&+& \left[ G^R_{a,b''} t_{b'',b} + G_{a,b'}^R t_{b',b} \right]
g_{b,b}^{+-} \left[ t_{b,b'} G_{b',a}^A + t_{b,b''} G_{b'',a}^A \right]\\
\label{terme3}
&+& G_{a,a}^R t_{a,a'} g_{a',a'}^{+-} t_{a',a} G_{a,a}^A\\
\label{terme4}
&+& G_{a,a}^R t_{a,a''} g_{a'',a''}^{+-} t_{a'',a} G_{a,a}^A\\
\label{terme5}
&+& G_{a,a}^R t_{a,a'} g_{a',b'}^{+-} t_{b',b} G_{b,a}^A\\
\label{terme6}
&+& G_{a,b}^R t_{b,b'} g_{b',a'}^{+-} t_{a',a} G_{a,a}^A\\
\label{terme7}
&+& G_{a,a}^R t_{a,a''} g_{a'',b''}^{+-} t_{b'',b} G_{b,a}^A\\
\label{terme8}
&+& G_{a,b}^R t_{b,b''} g_{b'',a''}^{+-} t_{a'',a} G_{a,a}^A\\
\label{terme9}
&+& G_{a,b}^R t_{b,b'} g_{b',b'}^{+-} t_{b',b} G_{b,a}^A\\
\label{terme10}
&+& G_{a,b}^R t_{b,b''} g_{b'',b''}^{+-} t_{b'',b} G_{b,a}^A
.
\end{eqnarray}
One finds
$G_{a,a}^{A,R}= \pm i \pi \tilde{\rho}_a$, and
$G_{a,a}^{+-} = 2 i \pi n_F(\omega) \tilde{\rho}_a$, with
the renormalized density of states
\begin{equation}
\label{eq:rho-ren-AB}
\tilde{\rho}_a = \frac{ \rho_a \left[
1 + \rho_b ( \gamma' + \gamma'') \right]}
{1 + (\rho_a + \rho_b) (\gamma' + \gamma'')
+ 2 \gamma' \gamma'' \rho_a \rho_b \left[1 -
\cos{\left( 2 \pi \phi / \phi_0 \right)} \right]}
,
\end{equation}
and
where I used the notation $\gamma'=\pi^2 t^2 \rho'$
and $\gamma'' = \pi^2 t^2 \rho''$ with
$t = |t_{a,a'}|=|t_{a,a''}|=|t_{b,b'}|
=|t_{b,b''}|$. 

Is is straightforward to use the same procedure
as in section~\ref{sec:calcul} to eliminate
disconnected processes, which leads to
$I^A_\beta = 
I^A_{\beta,{\rm local}} +
I^A_{\beta,\uparrow,{\rm non local}} +
I^A_{\beta,\downarrow,{\rm non local}} 
$, where the local contribution takes
the form
\begin{equation}
I^A_{\beta,{\rm local}} = 
4 \pi^2 |t_{x,\beta}|^4 \rho_{\beta,\uparrow}
\rho_{\beta,\downarrow} \int
d \omega \left[ n_F(\omega-eV) - n_F(\omega) \right]
|G_{x,x,1,2}^A|^2
,
\end{equation}
and the non local contribution is
\begin{eqnarray}
\nonumber
I^A_{\beta,\sigma,{\rm non local}} &=&
\int d \omega
\hat{Q} \left[
\frac{ 4 \pi^2 |t_{x,\beta}|^2 |t_{x,a}|^2
\rho_{\beta,\sigma} \rho_{a,-\sigma}^2
\rho_{b,-\sigma} \left[
(\gamma'_{-\sigma} + \gamma''_{-\sigma})^2
+2 \gamma'_{-\sigma} \gamma''_{-\sigma}
[1 - \cos{(2 \pi \phi / \phi_0)}] \right]}
{ \left[ 1 + (\rho_{a,-\sigma} + \rho_{b,-\sigma})
(\gamma'_{-\sigma} + \gamma''_{-\sigma})
+2 \gamma'_{-\sigma} \gamma''_{-\sigma}
\rho_{a,-\sigma} \rho_{b,-\sigma}
[1 - \cos{(2 \pi \phi / \phi_0)}] \right]^2} \right]\times\\
&& \left[ n_F(\omega-eV) - n_F(\omega) \right]
|G_{x,x,1,2}^A|^2
\label{eq:current-nl}
,
\end{eqnarray}
where the action of $\hat{Q}$ has already
been given.
The non local current Eq.~\ref{eq:current-nl} is
periodic in $\phi/\phi_0$. Such oscillations are 
a signature of non separable correlations.

The origin of these oscillations can be understood
in simple terms in the limit $t_{a,x}, t_{b,y} \ll 1$,
where the physics is the one of resonant tunneling.
There are discrete energy levels on the loop,
the energy of which depends on the magnetic
flux. Single electron
transmission is modulated
by the flux because the transmission is larger when
there is one resonant level.
Now the Andreev current is carried by correlated
pairs of electrons. If the Aharonov-Bohm loop
is off-resonance, one electron making the 
correlated pair (for instance a spin-up
electron) cannot be transmitted.
As a consequence, the whole correlated
pair cannot be transmitted. There is
therefore no spin-down current.

This situation is not far from the behavior
of the multichannel ferromagnet~--
superconductor junction. It is well known
that in such junctions there is no current
transmitted in the channels having only a 
spin-up Fermi surface~\cite{deJong}.

\begin{figure}
\centerline{\psfig{file=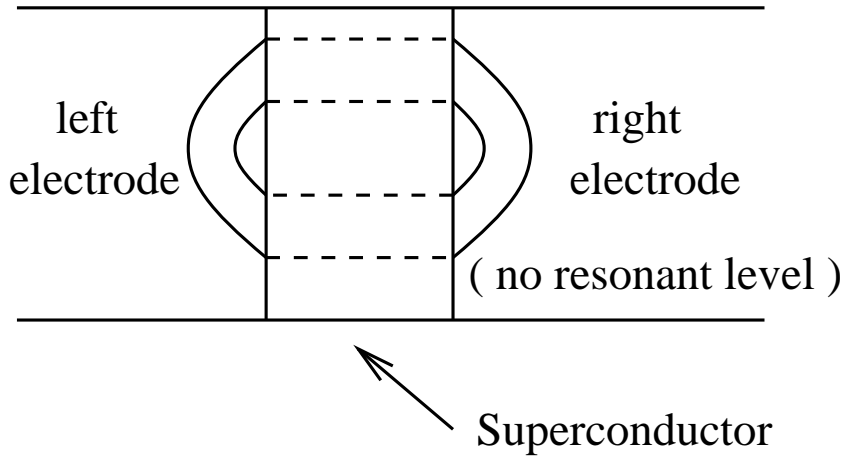,height=6cm}}
\caption{The diagram involved in the absence of
resonant tunneling in the right electrode.
This diagram does not carry electrical
charge.
\base
}
\label{fig:retro}
\end{figure}

The following equivalent explanation
can be found. There is one virtual state
in which one Cooper pair has been
extracted from the superconductor,
one electron is in the left electrode and
the other electron is in the right electrode.
Since there is no resonant level, the
electron in the right electrode is backscattered
onto the superconductor, undergoes an
Andreev reflection in which one Cooper
pair is transferred into the superconductor,
and a hole is transfered in the left electrode.
The whole processes is represented on Fig.~\ref{fig:retro}.
A quick look at this diagram shows that there
is no current transmitted. Note that the
processes involved in the multichannel
superconductor~-- ferromagnet junction can
be interpreted with the same diagram~\cite{Melin-2000}.

We stress that these two pictures
based on resonant tunneling hold
only if the
interfaces have low transparency contacts.
Nevertheless, since we found a systematic way to handle
higher order Feynman diagram, the model is valid also
for high transparency contacts,
in which case Aharonov-Bohm oscillations are also
predicted.

\section{Propagation of superconducting correlations
along a domain wall}
\label{sec:domain}
\begin{figure}
\centerline{\psfig{file=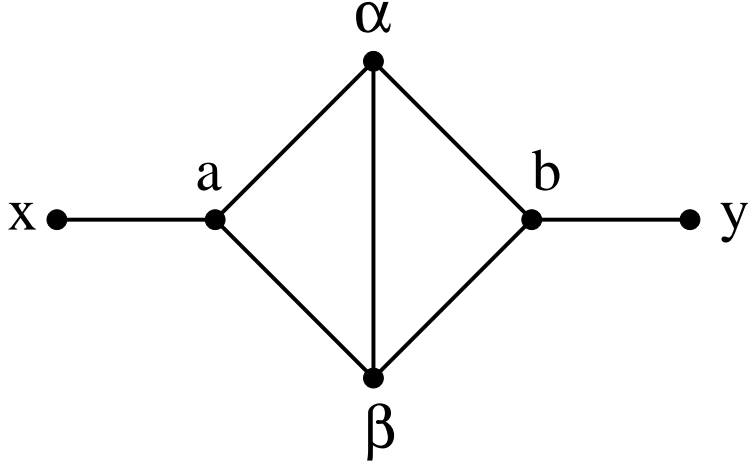,height=6cm}}
\caption{The circuit used to represent the propagation
of superconducting correlations along a domain
wall. There is a domain wall if $\alpha$ has
a spin-up magnetization and $\beta$ has a spin-down
magnetization. I note
$t=t_{a,\alpha}=t_{a,\beta}=t_{b,\alpha}=t_{b,\beta}$
and $t'=t_{\alpha,\beta}$. Site $x$ is superconducting.
\base
}
\label{fig:domain}
\end{figure}

Now, I illustrate the consequence
of the model regarding the propagation of superconducting
correlations along domain walls in ferromagnets.
Such propagation can generate
an enhancement of the proximity
effect in ferromagnet~/ superconductor
heterostructures, not against recent
experiments~\cite{Giroud,Petrashov,Chandra,Bauer}.
We use the geometry on Fig.~\ref{fig:domain}
to address this problem in an effective
single site formalism.
If site $\alpha$ has a spin-up
orientation and site $\beta$ has a spin-down orientation,
there is one domain wall in the junction. In the presence of
partially polarized ferromagnets, a spin--$\sigma$ electron
can be transferred across the domain wall because of the
hopping matrix element $t'$. This provides
a minimal model for Cooper pair
penetration along domain walls.
This model in which the domain wall has a vanishing
width
is expected to be relevant for describing a situation
in which the width of the domain wall is smaller
than the superconducting coherence length.

The Dyson equations corresponding to Fig.~\ref{fig:domain}
take the form
\begin{eqnarray}
G_{a,a} &=& g_{a,a} + g_{a,a} t_{a,\alpha}
G_{\alpha,a} + g_{a,a} t_{a,\beta} G_{\beta,a} \\
G_{\alpha,a} &=& g_{\alpha,\alpha} t_{\alpha,a} G_{a,a} +
g_{\alpha,\alpha} t_{\alpha,\beta} G_{\beta,a}
+ g_{\alpha,\alpha} T_{\alpha,b} G_{b,a}\\
G_{\beta,a} &=& g_{\beta,\beta} t_{\beta,a} G_{a,a}
+ g_{\beta,\beta} t_{\beta,\alpha} G_{\alpha,a}
+ g_{\beta,\beta} t_{\beta,b} G_{b,a}\\
G_{b,a} &=& g_{b,b} t_{b,\alpha} G_{\alpha,a} + g_{b,b} t_{b,\beta}
G_{\beta,a}
,
\end{eqnarray}
and similar equations hold for the Keldysh component.
These equations are solved into
$G_{a,a}^{A,R} = \pm i \pi \tilde{\rho}_a$,
$G^{+-}_{a,a} = 2 i \pi n_F(\omega)
\tilde{\rho}_a$, with the renormalized
density of states
\begin{equation}
\label{eq:rho-domain}
\tilde{\rho}_a = \rho_a
\frac{ 1 + \pi^2 t'^2 \rho_\alpha \rho_\beta
+ \pi^2 t^2 \rho_b  \left[ \rho_\alpha
+ \rho_\beta + 2 i \pi t' \rho_\alpha
\rho_\beta \right]}
{1 + \pi^2 t'^2 \rho_\alpha \rho_\beta 
+ \pi^2 t^2 (\rho_a + \rho_b)
\left[ \rho_\alpha + \rho_\beta
+2 i \pi t' \rho_\alpha \rho_\beta \right]}
.
\end{equation}
I deduce from Eq.~\ref{eq:rho-domain} that the
Andreev reflection current takes the form
\begin{eqnarray}
\label{eq:IA-domain1}
I^A &=& 4 \pi^6 t_{a,\alpha}^4 t^8
\int d \omega
{\hat{Q}} \left[
\prod_{\sigma=\uparrow,\downarrow}
\frac{ \rho_{x,\sigma} \rho_{a,\sigma}^2
\rho_{b,\sigma} \left[
(\rho_{\alpha,\sigma} + \rho_{\beta,\sigma})^2
+4 \pi^2 t'^2 \rho_{\alpha,\sigma}^2
\rho_{\beta,\sigma}^2 \right]}
{ \left| 1 + \pi^2 t'^2 \rho_{\alpha,\sigma}
\rho_{\beta,\sigma}
+ \pi^2 t^2 \left[ \rho_{\alpha,\sigma} +
\rho_{\beta,\sigma}
+2 i \pi t' \rho_{\alpha,\sigma}
\rho_{\beta,\sigma} \right] \right|^2} \right] \times\\
&& \left[ n_F(\omega-eV)
- n_F(\omega) \right] 
|G_{x,x,1,2}^A|^2
\label{eq:IA-domain2}
.
\end{eqnarray}
In the limit of fully polarized ferromagnets,
the current Eqs.~\ref{eq:IA-domain1},~\ref{eq:IA-domain2}
vanishes if the sites $\alpha$ and $\beta$ have
the same spin orientation. In the presence of
a domain wall, $\alpha$ and $\beta$ have an opposite
spin orientation, in which case Cooper
pairs can be transfered across the junction
even in the presence of fully polarized
ferromagnets.

The effect proposed here is related to two
other well known problems:
\begin{itemize}
\item[(i)] {\sl The $\pi$-junction:} It is well known
that Cooper pairs cannot propagate over large
distances in a ferromagnetic metal. In fact,
they cannot even propagate at all in a
fully polarized
single domain ferromagnet.
In partially
polarized ferromagnets, Cooper pair penetration
gives rise to the $\pi$-junction physics that
has been investigated by theorists since several
decades~\cite{Fulde,Larkin,Clogston,Demler},
and has been obtained experimentally only
recently~\cite{Ryazanov,Kontos}. We have shown
here that Cooper pairs can propagate in
fully spin polarized ferromagnets if there
are domain walls in the ferromagnet.
The number of conduction
channels is equal to the number of
domain walls.
\item[(ii)] {\sl Cryptomagnetism:} It has been
established a long time ago by Anderson and
Suhl that ferromagnetism and superconductivity
can accommodate each other in the same system if the
ferromagnet acquires a domain
structure~\cite{Anderson,Buzdin}. Our proposal
can be viewed as the proximity effect version
of cryptomagnetism.
\end{itemize}

\section{Conclusion}
\label{sec:conclusion}
To conclude, I have provided a minimal
formalism to treat non
separable correlations generated when a superconductor is
in contact with ferromagnetic or normal metal
electrodes.
It has been shown how to solve the problem of
disconnected contributions arising
in this formalism, and it has been given a systematic
way to handle the infinite series of Feynman diagram.

We answered the four questions given in
the introductory section, namely:
\begin{itemize}
\item[(i)] The superconducting gap depends
on the relative spin orientation of ferromagnetic
electrodes connected to a superconductor.
This might be probed in experiments.

\item[(ii)] It is possible to fabricate
linear superpositions of correlated
pairs of electrons. Transport in these systems
can be expressed in terms
of projections of the linear superposition
of correlated pairs of electrons. The 
wave function is projected on the relevant
quantities that participate to transport,
{\sl i.e.} the correlated pairs of electrons.

\item[(iii)] I proposed 
a new Aharonov-Bohm experiment to test non separable correlations.
In this experiment, the spin--$\sigma$ electron making the Cooper
is forced to couple to a magnetic flux while the
spin--($-\sigma$) electron has an ordinary propagation. It has been
shown that the spin--$\sigma$ current oscillates as a function
of the magnetic flux coupled to the spin-down electron.
It has been predicted that the effect appears
with low and high transparency contacts, and
independently of the spin polarization of the electrodes.
Therefore, normal metal electrodes can be used,
in which phase coherence can propagate
over large distances.

\item[(iv)] It has been shown that superconducting
correlations can propagate along ferromagnetic
domain walls. The model is strictly
speaking valid if
the width of the domain wall is small but
we are confident about forthcoming
generalizations of the model. The number
of Cooper pair conduction channels is equal to
the number of domain walls. This may apply
to existing experiments in
heterostructures~\cite{Giroud,Petrashov,Chandra}
which have not received a satisfactory
explanation up to now in spite of various
theoretical attempts~\cite{Chandra,Bauer}.
\end{itemize}

To end-up let us mention that the
work presented in this article calls both for
new experiments and for more sophisticated
theoretical descriptions.
It is in general needed to incorporate 
realistic ingredients in the models, which 
will be the subject of future publications.
Nevertheless, we have no doubt that the
same physics will be obtained in the presence
of realistic constraints.

\medskip

The author acknowledges fruitful discussions
with P. Degiovanni, D. Feinberg and M. Giroud.
This work is supported by the French Ministry
of Research under contract ACI 2086 CDR2.

\end{document}